\documentclass[10pt,twocolumn,aps,prl,superscriptaddress]{revtex4-2}

\usepackage[utf8]{inputenc}
\usepackage[T1]{fontenc}
\usepackage{amsmath,amssymb,graphicx}
\usepackage[colorlinks=true,linkcolor=blue,citecolor=blue,urlcolor=blue]{hyperref}

\begin{document}
\pagestyle{empty}

\title{Robust transfer of a quantum state from an absorbed photon into a diamond spin}

\author{Daisuke Ito}
\affiliation{Department of Physics, Graduate School of Engineering Science, Yokohama National University, 79-5, Tokiwadai, Hodogaya, Yokohama, 240-8501, Japan}

\author{Yuhei Sekiguchi}
\affiliation{Quantum Information Research Center, Institute of Advanced Sciences, Yokohama National University, 79-5, Tokiwadai, Hodogaya, Yokohama, 240-8501, Japan}

\author{Raustin Reyes}
\affiliation{Department of Physics, Graduate School of Engineering Science, Yokohama National University, 79-5, Tokiwadai, Hodogaya, Yokohama, 240-8501, Japan}

\author{Taichi Fujiwara}
\affiliation{Department of Physics, Graduate School of Engineering Science, Yokohama National University, 79-5, Tokiwadai, Hodogaya, Yokohama, 240-8501, Japan}

\author{Toshiharu Makino}
\affiliation{Quantum Information Research Center, Institute of Advanced Sciences, Yokohama National University, 79-5, Tokiwadai, Hodogaya, Yokohama, 240-8501, Japan}
\affiliation{Advanced Power Electronics Research Center, National Institute of Advanced Industrial Science and Technology, 1-1-1, Umezono, Tsukuba, Ibaraki, 305-8568, Japan}

\author{Hiromitsu Kato}
\affiliation{Quantum Information Research Center, Institute of Advanced Sciences, Yokohama National University, 79-5, Tokiwadai, Hodogaya, Yokohama, 240-8501, Japan}
\affiliation{Advanced Power Electronics Research Center, National Institute of Advanced Industrial Science and Technology, 1-1-1, Umezono, Tsukuba, Ibaraki, 305-8568, Japan}

\author{Hideo Kosaka}
\affiliation{Department of Physics, Graduate School of Engineering Science, Yokohama National University, 79-5, Tokiwadai, Hodogaya, Yokohama, 240-8501, Japan}
\affiliation{Quantum Information Research Center, Institute of Advanced Sciences, Yokohama National University, 79-5, Tokiwadai, Hodogaya, Yokohama, 240-8501, Japan}

\begin{abstract}
Conversion of a quantum state from a flying qubit to a memory qubit is crucial for distributed quantum computing. However, this requires precise spatiotemporal or frequency/phase alignment. Here, we experimentally demonstrate quantum teleportation-based state transfer from a photon into a spin in a nitrogen-vacancy center in diamond robust against both spectral and temporal errors. The achieved fidelity exceeds 0.94 within a frequency error of 100~MHz and 0.93 within an arrival-time error of 100~ns. This achievement enables extraordinarily robust entanglement generation between remote quantum memories compared with the conventional photon-interference-based approaches and paves the way for stable quantum networks.
\end{abstract}

\maketitle

Remote quantum entanglement serves as a fundamental resource for a wide range of applications, including quantum repeaters~\cite{Briegel_1998}, quantum sensing~\cite{Jozsa_2000}, and distributed quantum computing~\cite{Cirac_1999}. It has been experimentally realized using various physical platforms, such as neutral atoms~\cite{van_Leent_2022,Zhou_2024}, trapped ions~\cite{Krutyanskiy_2023}, quantum dots~\cite{Stockill_2017}, and diamond color centers~\cite{Knaut_2024}. Among these, the nitrogen-vacancy (NV) center in diamond has been widely used due to its long coherence time~\cite{Bradley_2019} and high controllability~\cite{Nagata_2018}, enabling remote entanglement generation~\cite{Hermans_2022,Bernien_2013}. To achieve entanglement in these platforms, photon-interference-based schemes, including the two-photon scheme~\cite{Barrett_2005,Hensen_2015} and the single-photon scheme~\cite{Cabrillo_1999,Duan_2001,Stolk_2024}, are widely used for remote entanglement generation. These schemes rely on precise mode matching of photons emitted independently from two distant nodes and interfered at a central beam splitter. Any mismatch in photonic modes, such as frequency, arrival time, or spatial mode, can degrade the entanglement fidelity. Additionally, the single-photon scheme requires phase locking between the nodes. To eliminate the requirement for mode matching, we focus on the quantum teleportation-based state transfer (QTST) scheme~\cite{Yang_2016,Tsurumoto_2019}, which stores an entangled photon state in the NV center. We experimentally demonstrate the robustness of the QTST against spectral and temporal mismatches.

\begin{figure}[t]
  \centering
  \includegraphics[width=0.9\linewidth]{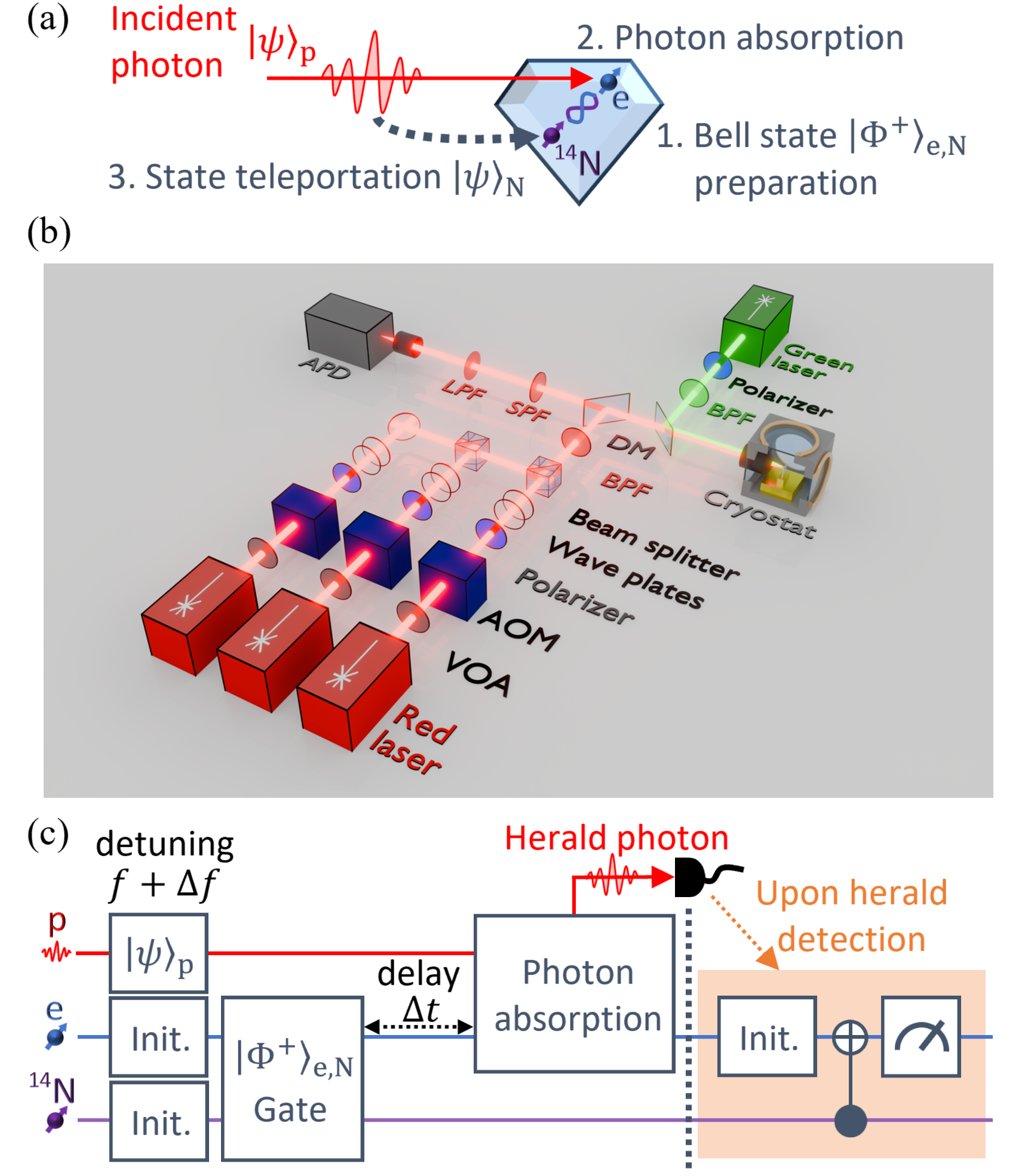}
  \caption{Quantum teleportation-based state transfer from a photon to a diamond NV center. (a) Schematic of the QTST protocol. (b) Optical setup. A  green laser passes through a bandpass filter (BPF) centered at $520~\mathrm{nm}$ with a $28~\mathrm{nm}$ bandwidth. Each red laser is assigned for electron spin initialization, readout, and incident photon preparation, and is filtered by a BPF centered at $637.21~\mathrm{nm}$ with a $0.5~\mathrm{nm}$ bandwidth. Acousto-optic modulators (AOM) and variable optical attenuators (VOA) are used to control the timing and intensity of the laser pulses. Heralding photons are detected by avalanche photodiodes (APD) after passing through a dichroic mirror (DM), a long-pass filter (LPF) with a cutoff at $645~\mathrm{nm}$, and a short-pass filter (SPF) with a cutoff at $765.5~\mathrm{nm}$. (c) Experimental sequence. An entangled state between the electron and nitrogen nuclear spins is first prepared, followed by photon absorption with controlled frequency and arrival-time errors. The nuclear spin is measured only upon successful herald detection.}
  \label{fig:fig1}
\end{figure}

The QTST scheme transfers the quantum state of an incident photon into a solid-state quantum memory via teleportation. The scheme first prepares an entangled state between the electron spin and the nitrogen nuclear spin of a diamond NV center, followed by absorption of the photon into a designated orbital excited state [Fig.~\ref{fig:fig1}(a)]. In this work, the spin Bell state
\begin{equation}
  \left|\Phi^{+}\right\rangle_{\mathrm{e,N}}
  =\frac{1}{\sqrt{2}}
   \bigl(\left|+1,+1\right\rangle_{\mathrm{e,N}}
        +\left|-1,-1\right\rangle_{\mathrm{e,N}}\bigr)
  \label{eq:bell_phi_plus}
\end{equation}
is prepared, where $\left|\mathrm{m_e}=0, \pm1\right\rangle_\mathrm{e}$ and $\left|\mathrm{m_N}=0, \pm1\right\rangle_\mathrm{N}$ denotes the electron and nitrogen nuclear spin states, and then an incident photon is absorbed only if the energy is resonant to the orbital excited state  $\left|A_{2}\right\rangle$ , which forms an entangled state between the orbital and spin states of the electron as described in Ref.~\cite{Maze_2011}
\begin{equation}
  \left|A_{2}\right\rangle
  = \frac{1}{\sqrt{2}}
    \bigl( \left|+1,-1\right\rangle_{\mathrm{p,e}}
         + \left|-1,+1\right\rangle_{\mathrm{p,e}} \bigr)
  = \left|\Psi^{+}\right\rangle_{\mathrm{p,e}}
  \label{eq:A2}
\end{equation}
where the orbital state is noted p since the orbital state of the electron and the polarization state of the photon coincides via the momentum selection rule \cite{Kosaka_2015}. Absorption into the $\left|A_{2}\right\rangle$ state thus acts as a probabilistic projection onto the Bell-basis state $\left|\Phi^{+}\right\rangle_{\mathrm{e,N}}$. The detection of a relaxation photon emitted from the $\left|A_{2}\right\rangle$ state serves as a herald, indicating that the incident photon has been absorbed and its quantum state has been teleported into the nuclear spin.

In the experimental demonstration, the polarization state of a red incident photon is transferred to the nitrogen nuclear spin in a diamond NV center through the following process. The NV center is first initialized to the negative charge state by a green laser. Three red lasers are individually assigned for distinct purposes: one for initialization of the electron spin, one for its readout~\cite{Robledo_2011}, and one for the preparation of the incident photon [Fig.~\ref{fig:fig1}(b)]. The incident photon is shaped into a $10~\mathrm{ns}$ pulse using an acousto-optic modulator (AOM), attenuated to an average absorption probability of 0.1 using a variable optical attenuator (VOA), and its polarization is adjusted via wave plates. To enhance optical efficiency, the NV is positioned in the center of the hemi-sphere curved solid immersion lens (SIL) with anti-reflection coating, and measured at $4~\mathrm{K}$. The ambient magnetic field is canceled using a triaxial Helmholtz coil. Microwave and radio-frequency pulses with arbitrary polarization are applied to control the electron and nuclear spins, using orthogonally wired antennas placed on the diamond near the SIL, as described in Ref.~\cite{Nagata_2018}. Relaxation photons emitted from the NV into the phonon sideband serve as heralding signals to announce the success of the quantum teleportation. These photons are spectrally filtered using a dichroic mirror in combination with short-pass and long-pass filters, followed by the detection with an avalanche photodiode (APD). The system is initialized into $\left|0,0\right\rangle_{\mathrm{e,N}}$ by applying laser pulses and spin control, and an entangled state between the electron and nuclear spins is subsequently generated via the hyperfine interaction~\cite{Nagata_2018} using a pulse sequence designed via gradient ascent pulse engineering (GRAPE)~\cite{Khaneja_2005} Incident photons subject to controlled arrival-time and frequency errors are then absorbed by this prepared state. The state of the nuclear spin is read out via the electron spin~\cite{kamimaki_2023} only when the heralding photon is detected [Fig.~\ref{fig:fig1}(c)].

\begin{figure}[t]
  \centering
  \includegraphics[width=0.9\linewidth]{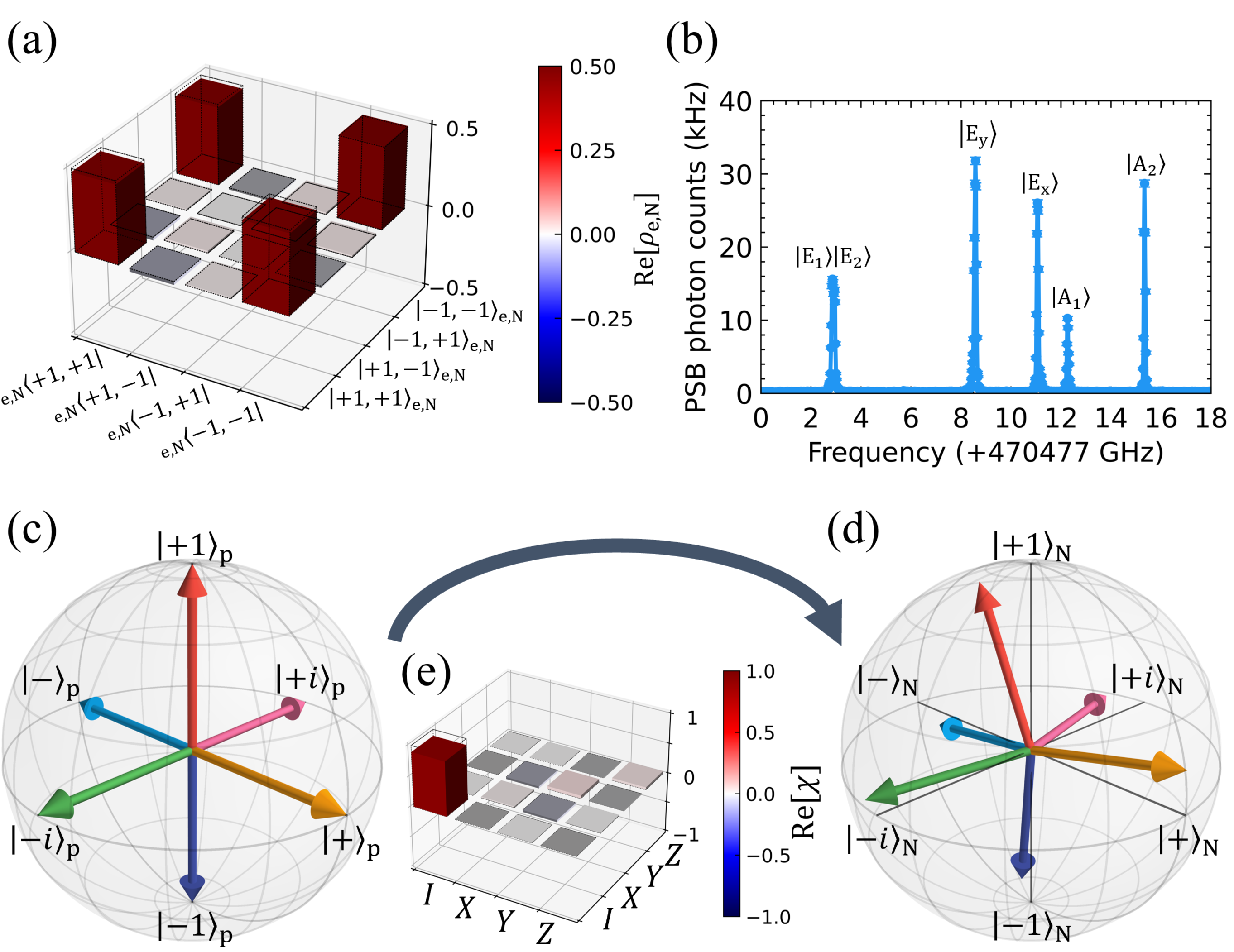}
  \caption{Characterization of entangled states and state transfer via QTST. (a) Quantum state tomography (QST) of the entangled state between the electron and nitrogen nuclear spins. The state fidelity is $0.97$. (b) Photoluminescence excitation (PLE) spectrum of the sample. The  $\left|E_{1}\right\rangle, \left|E_{2}\right\rangle$ states are degenerate orbital excited states corresponding to electron spin $\left|\pm1\right\rangle_\mathrm{e}$. Continuous resonant excitation to these states enables spin initialization into $\left|0\right\rangle_\mathrm{e}$. The $\left|E_{y}\right\rangle$ state is the orbital excited state for $\left|0\right\rangle_\mathrm{e}$ and is used for spin readout. The $\left|A_{2}\right\rangle$ state corresponds to the level defined in  Eq.~(\ref{eq:bell_phi_plus}). (c) Incident photon states. $\left|+1\right\rangle_\mathrm{p}$ and $\left|-1\right\rangle_\mathrm{p}$ correspond to right- and left-circular polarization, respectively; $\left|+\right\rangle_\mathrm{p}$ and $\left|-\right\rangle_\mathrm{p}$ to horizontal and vertical linear polarization; and $\left|+i\right\rangle_\mathrm{p}$ and $\left|-i\right\rangle_\mathrm{p}$ to diagonal and anti-diagonal linear polarization. (d) QST of the nuclear spin after state transfer. The average fidelity over six input states is $0.94$. (e) Real part of the process $\chi$ matrix reconstructed via quantum process tomography (QPT). Ideally, only the identity–identity (II) component is nonzero.}
  \label{fig:fig2}
\end{figure}

The first step of the QTST scheme is the preparation of a quantum entangled state between the electron and nitrogen nuclear spins. The density matrix reconstructed by the quantum state tomography (QST) is shown in Fig.~\ref{fig:fig2}(a), which reveals the state fidelity to be $F=0.97$. The resonance frequency of the orbital excited state is identified via photoluminescence excitation (PLE) spectroscopy [Fig.~\ref{fig:fig2}(b)], which allows extraction of the crystal strain of $\delta_\perp = 1.25~\mathrm{GHz}$ along the direction orthogonal to the NV axis, estimated by the excited-state level splitting~\cite{Batalov_2009}. After the entangled state in Eq.~\ref{eq:bell_phi_plus} is prepared, incident photons in the six polarization states shown in Fig.~\ref{fig:fig2}(c) are absorbed into the electron spin, and their quantum states are transferred to the nitrogen nuclear spin via the QTST scheme. The reconstructed density matrix of the nuclear spin is shown in Fig.~\ref{fig:fig2}(d). The corresponding $\chi$ matrix of the QTST process, reconstructed via quantum process tomography (QPT), is also shown in Fig.~\ref{fig:fig2}(e). The fidelity of the transferred states, averaged over the six inputs without frequency or arrival-time error, is estimated to be $F=0.94$.

\begin{figure}[t]
  \centering
  \includegraphics[width=0.9\linewidth]{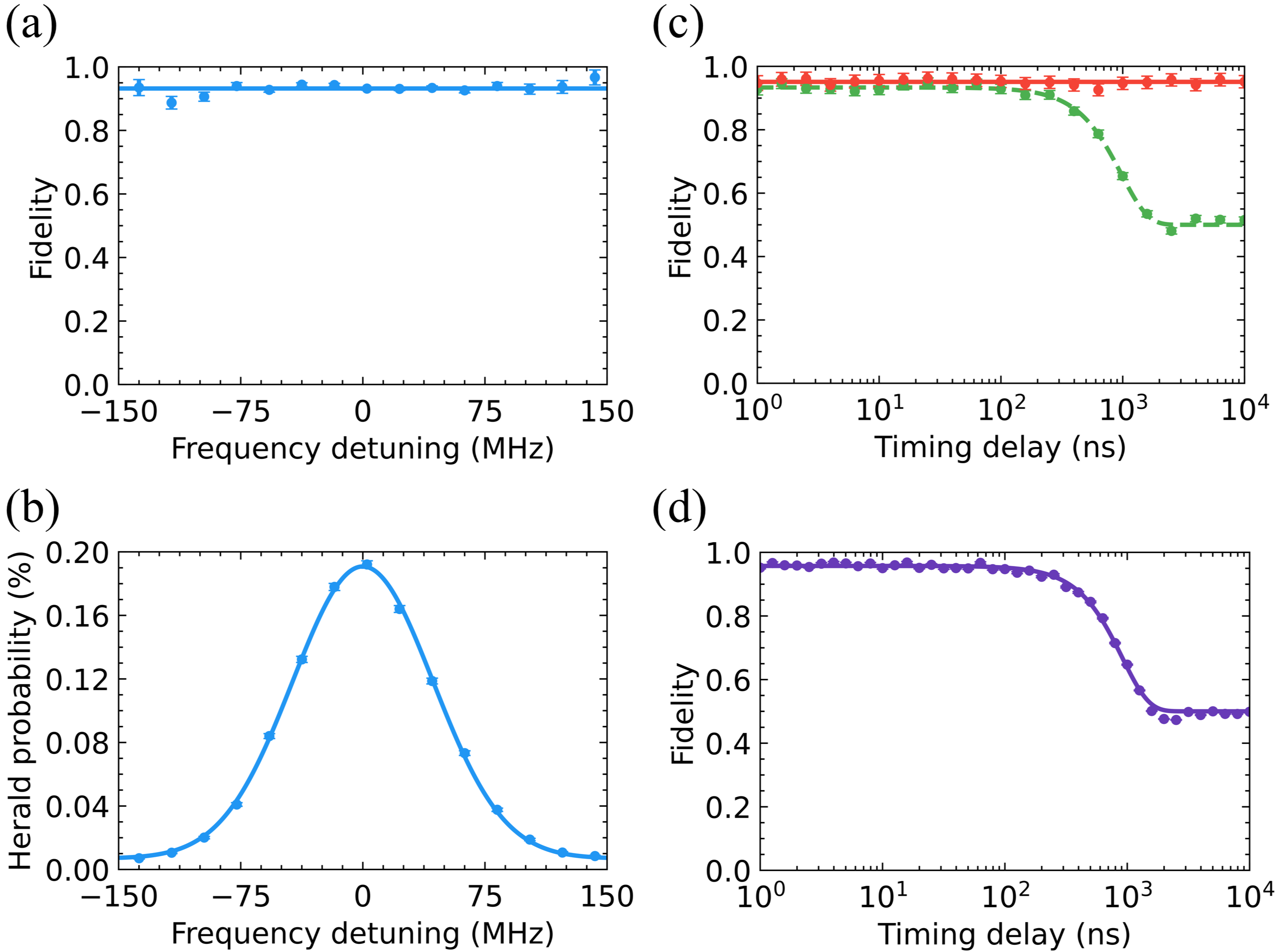}
  \caption{Experimental evaluation of the QTST protocol under frequency and arrival-time errors. (a) Fidelity under frequency errors. The solid line indicates a constant fit. (b) Heralding probability under frequency errors with an average of $0.1$ incident photons. (c) Fidelity under arrival-time errors. Orange points show the average fidelity for input states$\left|+1\right\rangle_\mathrm{p}$ and $\left|-1\right\rangle_\mathrm{p}$ with a solid-line fit. Green points show the average fidelity for $\left|+\right\rangle_\mathrm{p}, \left|-\right\rangle_\mathrm{p}, \left|+i\right\rangle_\mathrm{p}$ and $\left|-i\right\rangle_\mathrm{p}$ with a dashed-line fit. (d) Fidelity of the electron–nuclear spin entangled state. Error bars represent standard deviations derived from photon shot noise.}
  \label{fig:fig3}
\end{figure}

The robustness of the QTST scheme against frequency errors is evaluated by the fidelity degradation as a function of frequency detuning of the input photon from the $\left|A_2\right\rangle$ resonance. The average fidelity remains at $F=0.94$ over the detuning of $100~\mathrm{MHz}$ [Fig.~\ref{fig:fig3}(a)], indicating that the fidelity is unaffected by frequency errors although the heralding probability decreases [Fig.~\ref{fig:fig3}(b)]. Heralding events are observed across a frequency range (standard deviation: $61~\mathrm{MHz}$) over the natural linewidth due to spectral diffusion. On the other hand, the robustness against arrival-time errors is estimated by the fidelity as a function of the timing delay of the photon absorption from the preparation of the entangled state in Eq.~\ref{eq:bell_phi_plus} [Fig.~\ref{fig:fig3}(c)]. Different behaviors are observed depending on the input polarization state of the incident photon: the fidelity remains the same for $\left|+1\right\rangle_\mathrm{p}$ and $\left|-1\right\rangle_\mathrm{p}$, whereas it changes for $\left|+\right\rangle_\mathrm{p}$, $\left|-\right\rangle_\mathrm{p}$, $\left|+i\right\rangle_\mathrm{p}$, and $\left|-i\right\rangle_\mathrm{p}$. The standard deviation is $0.91~\mu\mathrm{s}$, indicating robustness against realistic timing jitter below $1~\mathrm{ns}$ and temporal drifts on the order of tens of nanoseconds~\cite{Bersin_2024}.

\begin{figure}[t]
  \centering
  \includegraphics[width=0.9\linewidth]{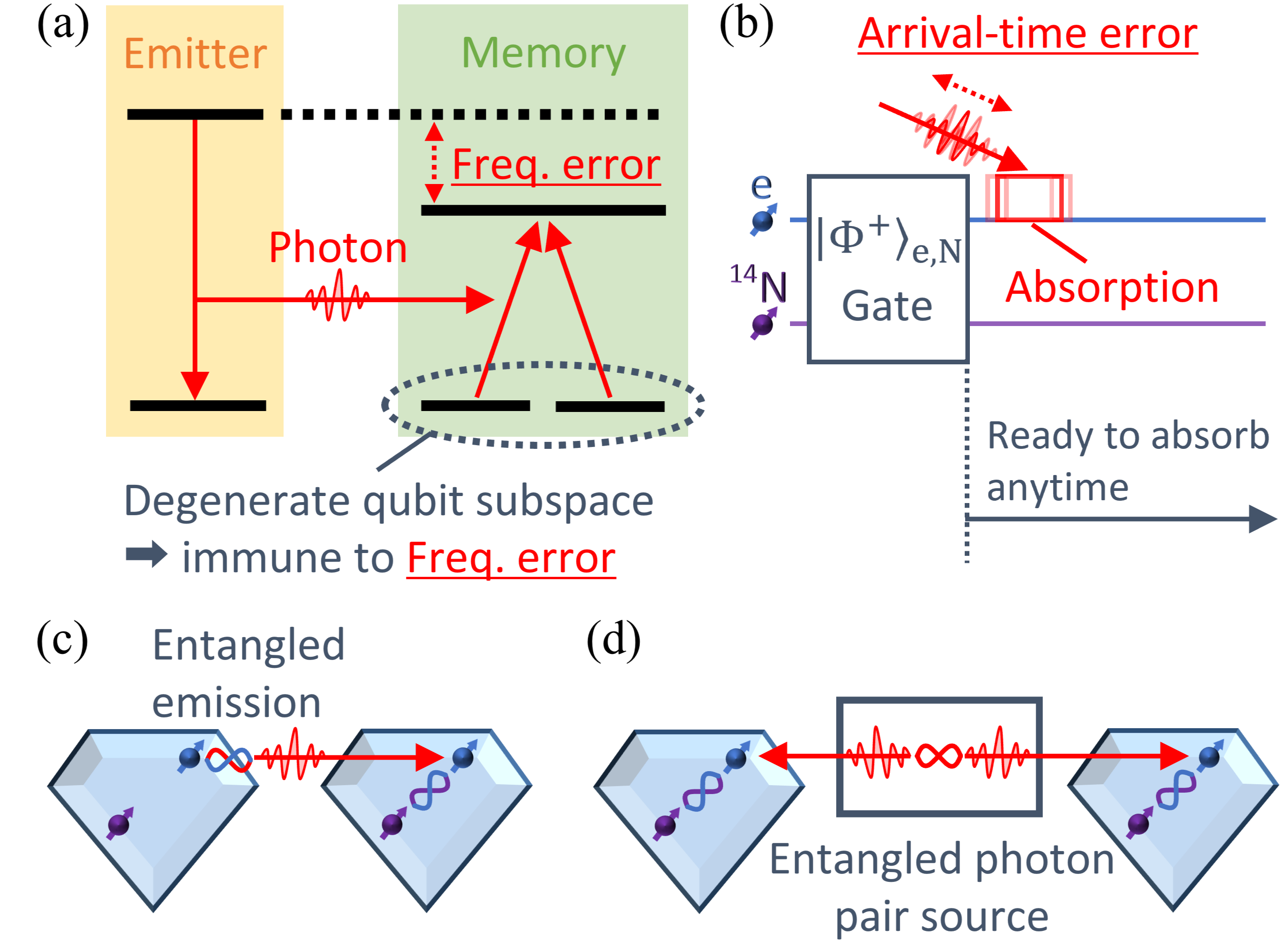}
  \caption{Robustness and applications of the QTST scheme. (a) Conceptual illustration of QTST under frequency errors. Frequency detuning does not degrade the fidelity of QTST. (b) Conceptual illustration of QTST under arrival-time errors. Once the entangled state between the electron and nuclear spins is prepared, the incident photon can ideally arrive at any time without affecting the fidelity. (c) Remote entanglement generation via QTST with an entangled photon emitter. (d) Remote entanglement generation via QTST with an entangled photon pair source.}
  \label{fig:fig4}
\end{figure}

The QTST scheme is robust against the frequency detuning, which does not affect the degenerate spin qubit, since the heralding signal guarantees the success of Bell-state measurement even in the presence of the detuning as shown in Fig.~\ref{fig:fig4}(a). The robustness remains working even with inhomogeneous broadening due to photoinduced charge fluctuations~\cite{McCullian_2022} and regardless of the photon arrival time once the entangled state between the electron and nuclear spins has been prepared as shown in Fig.~\ref{fig:fig4}(b).

The QTST scheme can also be applied to remote entanglement generation as in the following two examples. A photon entangled with one NV spin is transferred to the remote NV spin~\cite{Sekiguchi_2021} [Fig.~\ref{fig:fig4}(c)], or entangled photons are distributed and transferred to separate NV spins, generating a remote entanglement [Fig.~\ref{fig:fig4}(d)]. The fidelity of the remote entangled state remains unaffected by spectral and timing errors as the transferring process is robust against those errors.

The origin of the fidelity degradation observed under arrival-time errors is phase relaxation of the electron--nitrogen entangled state due to the multitude of weakly coupled $^{13}\mathrm{C}$ nuclear spins surrounding the NV center. Immediately after preparing the state of Eq.~\ref{eq:bell_phi_plus} the density matrix is
\begin{equation}
  \rho_{e,N}(t=0)=\lvert\Phi^{+}\rangle_{e,N}\langle\Phi^{+}\rvert
  \label{eq:rho_initial}
\end{equation}
In the presence of many weakly coupled  $^{13}\mathrm{C}$ spins, however, the off-diagonal coherence decays, and the state evolves towards
\begin{align}
  \rho_{e,N}(t\!\to\!\infty) 
  = \frac{\lvert\Phi^{+}\rangle_{e,N}\langle\Phi^{+}\rvert
        + \lvert\Phi^{-}\rangle_{e,N}\langle\Phi^{-}\rvert}{2},
  \label{eq:rho_longtime}
\end{align}
as phase decoherence leads to a mixture of Bell states. If an arbitrary photonic state
\begin{equation}
  \lvert\psi\rangle_{p}= \alpha\,\lvert +1\rangle_{p}
                         + \beta\,\lvert -1\rangle_{p}
  \label{eq:psi_photon}
\end{equation}
is teleported via the Bell-state measurement in Eq.~(\ref{eq:A2}),  
the nuclear-spin state obtained after the $\sigma_x$ feed-forward becomes
\begin{equation}
  \rho_{N} = \alpha^{2}\lvert +1\rangle_{N}\langle +1\rvert
           + \beta^{2}\lvert -1\rangle_{N}\langle -1\rvert.
  \label{eq:rho_nuclear}
\end{equation}
This means that only the computational-basis states $\left|+1\right\rangle_\mathrm{N} \langle+1|$ and $\left|-1\right\rangle_\mathrm{N} \langle-1|$ are faithfully transferred with large arrival-time errors, whereas all other superposition states are mapped onto mixed states. The fidelity of the electron--nuclear entangled state as a function of the delay between the entanglement preparation and the Bell-state projection, shown in Fig.~\ref{fig:fig3}(d), relaxes towards the classical 0.5 limit predicted by Eq.~\ref{eq:rho_longtime}. The standard deviation, $0.98~\mu\mathrm{s}$, is consistent with the fidelity decay observed for the input states $\left|+\right\rangle_\mathrm{p}$, $\left|-\right\rangle_\mathrm{p}$, $\left|+i\right\rangle_\mathrm{p}$, and $\left|-i\right\rangle_\mathrm{p}$. The diamond used in the demonstration contains $1.1\%~^{13}\mathrm{C}$ at natural abundance. The robustness against arrival-time errors can be further enhanced by using isotopically purified diamond or applying dynamical decoupling techniques.

The degradation in fidelity also originates from the state preparation and measurement (SPAM) errors and strain-induced eigenstate change in the excited state. The SPAM errors arise from imperfect orthogonality of the crossed microwave antennae and inaccuracies in the applied microwave pulses, and can be mitigated through device improvements or pre-compensation. The residual population on $\left|0\right\rangle_\mathrm{N}$ is identified as SPAM errors since the transferred quantum state should be in the nuclear-spin qubit subspace spanned by $\left|\pm1\right\rangle_\mathrm{N}$. The probability of SPAM errors in the demonstration is $1.6\%$, which will be post-selectively eliminated. When the QTST scheme is used for remote entanglement generation, the SPAM error can be handled non-destructively: measurement of $\left|0\right\rangle_\mathrm{N}$ without disturbing $\left|\pm1\right\rangle_\mathrm{N}$ allows the heralding event to be discarded, enabling the scheme to be retried and preventing any fidelity loss. The influence of crystal strain manifests as a deviation of the excited state $\left|A_2\right\rangle$ from the ideal eigenstate described in Eq.~\ref{eq:A2}. The fidelity between the eigenvector of the strain-influenced Hamiltonian—estimated from the PLE spectrum—and the ideal strain-free eigenstate $\left|\Psi^{+}\right\rangle_{p,e}$ is estimated to be $F = 0.98$. This strain effect can be compensated by applying a static electric field~\cite{Bassett_2011}, offering a viable method to further enhance the fidelity of the QTST scheme.

The QTST scheme enables robust entanglement generation rather than interference-based approaches. Although the robustness of interference-based schemes can be partially improved by using narrowband frequency filters, fidelity degradation still arises due to frequency mismatches within the finite filter bandwidth. In contrast, the absorption spectrum in the QTST scheme functions as a frequency filter that passively tracks spectral diffusion and does not induce fidelity degradation. This robustness eliminates the need to stabilize spectral and temporal degrees of freedom, leaving only polarization in long-distance quantum channels such as deployed optical fibers, thereby simplifying system design. In addition, since the QTST scheme does not induce fidelity degradation due to mode mismatch, it is tolerant to device inhomogeneity and enables entanglement generation even when using entangled-photon pair sources or entirely different physical platforms, such as neutral atoms. Robustness against arrival-time errors further facilitates temporal multiplexing, thereby enhancing entanglement generation rates.

While the QTST scheme exhibits high robustness, its entanglement-generation rate scales similarly to that of two-photon schemes and remains lower than the rates achievable with one-photon schemes. For a channel transmittance $\eta$ between two nodes separated by a distance $L$, one-photon schemes succeed with a probability proportional to $\sqrt{\eta}$, whereas two-photon schemes and those illustrated in Figs.~\ref{fig:fig4}(c) and \ref{fig:fig4}(d) scale linearly with $\eta$~\cite{Sangouard_2011}. Furthermore, the dependence on the zero-phonon line (ZPL) emission fraction differs: it scales linearly for the one-photon scheme, but quadratically for the others. However, the ZPL emission fraction can be effectively enhanced by employing photonic crystal structures via the Purcell effect~\cite{Jung_2019}. In addition, the rate gap can be reduced under low-loss conditions.

We have experimentally demonstrated the robustness of the QTST scheme. The demonstration has revealed that QTST-based entanglement generation schemes provide significantly more robust than interference-based schemes with high fidelity and minimal error correction, offering a scalable and practical approach for distributed quantum computing and multi-node quantum repeater architecture.

\begin{acknowledgments}
\textit{Acknowledgments}—H. Kosaka acknowledges the funding support from Japan Science and Technology Agency (JST) Moonshot R\&Dgrant (JPMJMS2062) and JST CREST grant (JPMJCR1773). H. Kosaka also acknowledges the Ministry of Internal Affairs and Communications (MIC) for funding, R\&D for construction of global quantum cryptography network (JPMI00316), R\&D of ICT Priority Technology Project, and the Japan Society for the Promotion of Science (JSPS) Grants-in-Aid for Scientific Research (20H05661, 20K20441, 25H0083050).
\end{acknowledgments}


\begin{thebibliography}{31}%
\makeatletter
\providecommand \@ifxundefined [1]{%
 \@ifx{#1\undefined}
}%
\providecommand \@ifnum [1]{%
 \ifnum #1\expandafter \@firstoftwo
 \else \expandafter \@secondoftwo
 \fi
}%
\providecommand \@ifx [1]{%
 \ifx #1\expandafter \@firstoftwo
 \else \expandafter \@secondoftwo
 \fi
}%
\providecommand \natexlab [1]{#1}%
\providecommand \enquote  [1]{``#1''}%
\providecommand \bibnamefont  [1]{#1}%
\providecommand \bibfnamefont [1]{#1}%
\providecommand \citenamefont [1]{#1}%
\providecommand \href@noop [0]{\@secondoftwo}%
\providecommand \href [0]{\begingroup \@sanitize@url \@href}%
\providecommand \@href[1]{\@@startlink{#1}\@@href}%
\providecommand \@@href[1]{\endgroup#1\@@endlink}%
\providecommand \@sanitize@url [0]{\catcode `\\12\catcode `\$12\catcode `\&12\catcode `\#12\catcode `\^12\catcode `\_12\catcode `\%12\relax}%
\providecommand \@@startlink[1]{}%
\providecommand \@@endlink[0]{}%
\providecommand \url  [0]{\begingroup\@sanitize@url \@url }%
\providecommand \@url [1]{\endgroup\@href {#1}{\urlprefix }}%
\providecommand \urlprefix  [0]{URL }%
\providecommand \Eprint [0]{\href }%
\providecommand \doibase [0]{https://doi.org/}%
\providecommand \selectlanguage [0]{\@gobble}%
\providecommand \bibinfo  [0]{\@secondoftwo}%
\providecommand \bibfield  [0]{\@secondoftwo}%
\providecommand \translation [1]{[#1]}%
\providecommand \BibitemOpen [0]{}%
\providecommand \bibitemStop [0]{}%
\providecommand \bibitemNoStop [0]{.\EOS\space}%
\providecommand \EOS [0]{\spacefactor3000\relax}%
\providecommand \BibitemShut  [1]{\csname bibitem#1\endcsname}%
\let\auto@bib@innerbib\@empty
\bibitem [{\citenamefont {Briegel}\ \emph {et~al.}(1998)\citenamefont {Briegel}, \citenamefont {Dür}, \citenamefont {Cirac},\ and\ \citenamefont {Zoller}}]{Briegel_1998}%
  \BibitemOpen
  \bibfield  {author} {\bibinfo {author} {\bibfnamefont {H.-J.}\ \bibnamefont {Briegel}}, \bibinfo {author} {\bibfnamefont {W.}~\bibnamefont {Dür}}, \bibinfo {author} {\bibfnamefont {J.~I.}\ \bibnamefont {Cirac}},\ and\ \bibinfo {author} {\bibfnamefont {P.}~\bibnamefont {Zoller}},\ }\href {http://dx.doi.org/10.1103/PhysRevLett.81.5932} {\bibfield  {journal} {\bibinfo  {journal} {Physical Review Letters}\ }\textbf {\bibinfo {volume} {81}},\ \bibinfo {pages} {5932–5935} (\bibinfo {year} {1998})}\BibitemShut {NoStop}%
\bibitem [{\citenamefont {Jozsa}\ \emph {et~al.}(2000)\citenamefont {Jozsa}, \citenamefont {Abrams}, \citenamefont {Dowling},\ and\ \citenamefont {Williams}}]{Jozsa_2000}%
  \BibitemOpen
  \bibfield  {author} {\bibinfo {author} {\bibfnamefont {R.}~\bibnamefont {Jozsa}}, \bibinfo {author} {\bibfnamefont {D.~S.}\ \bibnamefont {Abrams}}, \bibinfo {author} {\bibfnamefont {J.~P.}\ \bibnamefont {Dowling}},\ and\ \bibinfo {author} {\bibfnamefont {C.~P.}\ \bibnamefont {Williams}},\ }\href {http://dx.doi.org/10.1103/PhysRevLett.85.2010} {\bibfield  {journal} {\bibinfo  {journal} {Physical Review Letters}\ }\textbf {\bibinfo {volume} {85}},\ \bibinfo {pages} {2010} (\bibinfo {year} {2000})}\BibitemShut {NoStop}%
\bibitem [{\citenamefont {Cirac}\ \emph {et~al.}(1999)\citenamefont {Cirac}, \citenamefont {Ekert}, \citenamefont {Huelga},\ and\ \citenamefont {Macchiavello}}]{Cirac_1999}%
  \BibitemOpen
  \bibfield  {author} {\bibinfo {author} {\bibfnamefont {J.~I.}\ \bibnamefont {Cirac}}, \bibinfo {author} {\bibfnamefont {A.~K.}\ \bibnamefont {Ekert}}, \bibinfo {author} {\bibfnamefont {S.~F.}\ \bibnamefont {Huelga}},\ and\ \bibinfo {author} {\bibfnamefont {C.}~\bibnamefont {Macchiavello}},\ }\href {http://dx.doi.org/10.1103/PhysRevA.59.4249} {\bibfield  {journal} {\bibinfo  {journal} {Physical Review A}\ }\textbf {\bibinfo {volume} {59}},\ \bibinfo {pages} {4249} (\bibinfo {year} {1999})}\BibitemShut {NoStop}%
\bibitem [{\citenamefont {van Leent}\ \emph {et~al.}(2022)\citenamefont {van Leent}, \citenamefont {Bock}, \citenamefont {Fertig}, \citenamefont {Garthoff}, \citenamefont {Eppelt}, \citenamefont {Zhou}, \citenamefont {Malik}, \citenamefont {Seubert}, \citenamefont {Bauer}, \citenamefont {Rosenfeld}, \citenamefont {Zhang}, \citenamefont {Becher},\ and\ \citenamefont {Weinfurter}}]{van_Leent_2022}%
  \BibitemOpen
  \bibfield  {author} {\bibinfo {author} {\bibfnamefont {T.}~\bibnamefont {van Leent}}, \bibinfo {author} {\bibfnamefont {M.}~\bibnamefont {Bock}}, \bibinfo {author} {\bibfnamefont {F.}~\bibnamefont {Fertig}}, \bibinfo {author} {\bibfnamefont {R.}~\bibnamefont {Garthoff}}, \bibinfo {author} {\bibfnamefont {S.}~\bibnamefont {Eppelt}}, \bibinfo {author} {\bibfnamefont {Y.}~\bibnamefont {Zhou}}, \bibinfo {author} {\bibfnamefont {P.}~\bibnamefont {Malik}}, \bibinfo {author} {\bibfnamefont {M.}~\bibnamefont {Seubert}}, \bibinfo {author} {\bibfnamefont {T.}~\bibnamefont {Bauer}}, \bibinfo {author} {\bibfnamefont {W.}~\bibnamefont {Rosenfeld}}, \bibinfo {author} {\bibfnamefont {W.}~\bibnamefont {Zhang}}, \bibinfo {author} {\bibfnamefont {C.}~\bibnamefont {Becher}},\ and\ \bibinfo {author} {\bibfnamefont {H.}~\bibnamefont {Weinfurter}},\ }\href {http://dx.doi.org/10.1038/s41586-022-04764-4} {\bibfield  {journal} {\bibinfo  {journal} {Nature}\ }\textbf {\bibinfo {volume} {607}},\ \bibinfo {pages} {69–73} (\bibinfo
  {year} {2022})}\BibitemShut {NoStop}%
\bibitem [{\citenamefont {Zhou}\ \emph {et~al.}(2024)\citenamefont {Zhou}, \citenamefont {Malik}, \citenamefont {Fertig}, \citenamefont {Bock}, \citenamefont {Bauer}, \citenamefont {van Leent}, \citenamefont {Zhang}, \citenamefont {Becher},\ and\ \citenamefont {Weinfurter}}]{Zhou_2024}%
  \BibitemOpen
  \bibfield  {author} {\bibinfo {author} {\bibfnamefont {Y.}~\bibnamefont {Zhou}}, \bibinfo {author} {\bibfnamefont {P.}~\bibnamefont {Malik}}, \bibinfo {author} {\bibfnamefont {F.}~\bibnamefont {Fertig}}, \bibinfo {author} {\bibfnamefont {M.}~\bibnamefont {Bock}}, \bibinfo {author} {\bibfnamefont {T.}~\bibnamefont {Bauer}}, \bibinfo {author} {\bibfnamefont {T.}~\bibnamefont {van Leent}}, \bibinfo {author} {\bibfnamefont {W.}~\bibnamefont {Zhang}}, \bibinfo {author} {\bibfnamefont {C.}~\bibnamefont {Becher}},\ and\ \bibinfo {author} {\bibfnamefont {H.}~\bibnamefont {Weinfurter}},\ }\href {http://dx.doi.org/10.1103/PRXQuantum.5.020307} {\bibfield  {journal} {\bibinfo  {journal} {PRX Quantum}\ }\textbf {\bibinfo {volume} {5}} (\bibinfo {year} {2024})}\BibitemShut {NoStop}%
\bibitem [{\citenamefont {Krutyanskiy}\ \emph {et~al.}(2023)\citenamefont {Krutyanskiy}, \citenamefont {Galli}, \citenamefont {Krcmarsky}, \citenamefont {Baier}, \citenamefont {Fioretto}, \citenamefont {Pu}, \citenamefont {Mazloom}, \citenamefont {Sekatski}, \citenamefont {Canteri}, \citenamefont {Teller}, \citenamefont {Schupp}, \citenamefont {Bate}, \citenamefont {Meraner}, \citenamefont {Sangouard}, \citenamefont {Lanyon},\ and\ \citenamefont {Northup}}]{Krutyanskiy_2023}%
  \BibitemOpen
  \bibfield  {author} {\bibinfo {author} {\bibfnamefont {V.}~\bibnamefont {Krutyanskiy}}, \bibinfo {author} {\bibfnamefont {M.}~\bibnamefont {Galli}}, \bibinfo {author} {\bibfnamefont {V.}~\bibnamefont {Krcmarsky}}, \bibinfo {author} {\bibfnamefont {S.}~\bibnamefont {Baier}}, \bibinfo {author} {\bibfnamefont {D.~A.}\ \bibnamefont {Fioretto}}, \bibinfo {author} {\bibfnamefont {Y.}~\bibnamefont {Pu}}, \bibinfo {author} {\bibfnamefont {A.}~\bibnamefont {Mazloom}}, \bibinfo {author} {\bibfnamefont {P.}~\bibnamefont {Sekatski}}, \bibinfo {author} {\bibfnamefont {M.}~\bibnamefont {Canteri}}, \bibinfo {author} {\bibfnamefont {M.}~\bibnamefont {Teller}}, \bibinfo {author} {\bibfnamefont {J.}~\bibnamefont {Schupp}}, \bibinfo {author} {\bibfnamefont {J.}~\bibnamefont {Bate}}, \bibinfo {author} {\bibfnamefont {M.}~\bibnamefont {Meraner}}, \bibinfo {author} {\bibfnamefont {N.}~\bibnamefont {Sangouard}}, \bibinfo {author} {\bibfnamefont {B.~P.}\ \bibnamefont {Lanyon}},\ and\ \bibinfo {author} {\bibfnamefont {T.~E.}\
  \bibnamefont {Northup}},\ }\href {http://dx.doi.org/10.1103/PhysRevLett.130.050803} {\bibfield  {journal} {\bibinfo  {journal} {Physical Review Letters}\ }\textbf {\bibinfo {volume} {130}} (\bibinfo {year} {2023})}\BibitemShut {NoStop}%
\bibitem [{\citenamefont {Stockill}\ \emph {et~al.}(2017)\citenamefont {Stockill}, \citenamefont {Stanley}, \citenamefont {Huthmacher}, \citenamefont {Clarke}, \citenamefont {Hugues}, \citenamefont {Miller}, \citenamefont {Matthiesen}, \citenamefont {Le~Gall},\ and\ \citenamefont {Atatüre}}]{Stockill_2017}%
  \BibitemOpen
  \bibfield  {author} {\bibinfo {author} {\bibfnamefont {R.}~\bibnamefont {Stockill}}, \bibinfo {author} {\bibfnamefont {M.}~\bibnamefont {Stanley}}, \bibinfo {author} {\bibfnamefont {L.}~\bibnamefont {Huthmacher}}, \bibinfo {author} {\bibfnamefont {E.}~\bibnamefont {Clarke}}, \bibinfo {author} {\bibfnamefont {M.}~\bibnamefont {Hugues}}, \bibinfo {author} {\bibfnamefont {A.}~\bibnamefont {Miller}}, \bibinfo {author} {\bibfnamefont {C.}~\bibnamefont {Matthiesen}}, \bibinfo {author} {\bibfnamefont {C.}~\bibnamefont {Le~Gall}},\ and\ \bibinfo {author} {\bibfnamefont {M.}~\bibnamefont {Atatüre}},\ }\href {http://dx.doi.org/10.1103/PhysRevLett.119.010503} {\bibfield  {journal} {\bibinfo  {journal} {Physical Review Letters}\ }\textbf {\bibinfo {volume} {119}} (\bibinfo {year} {2017})}\BibitemShut {NoStop}%
\bibitem [{\citenamefont {Knaut}\ \emph {et~al.}(2024)\citenamefont {Knaut}, \citenamefont {Suleymanzade}, \citenamefont {Wei}, \citenamefont {Assumpcao}, \citenamefont {Stas}, \citenamefont {Huan}, \citenamefont {Machielse}, \citenamefont {Knall}, \citenamefont {Sutula}, \citenamefont {Baranes}, \citenamefont {Sinclair}, \citenamefont {De-Eknamkul}, \citenamefont {Levonian}, \citenamefont {Bhaskar}, \citenamefont {Park}, \citenamefont {Lončar},\ and\ \citenamefont {Lukin}}]{Knaut_2024}%
  \BibitemOpen
  \bibfield  {author} {\bibinfo {author} {\bibfnamefont {C.~M.}\ \bibnamefont {Knaut}}, \bibinfo {author} {\bibfnamefont {A.}~\bibnamefont {Suleymanzade}}, \bibinfo {author} {\bibfnamefont {Y.-C.}\ \bibnamefont {Wei}}, \bibinfo {author} {\bibfnamefont {D.~R.}\ \bibnamefont {Assumpcao}}, \bibinfo {author} {\bibfnamefont {P.-J.}\ \bibnamefont {Stas}}, \bibinfo {author} {\bibfnamefont {Y.~Q.}\ \bibnamefont {Huan}}, \bibinfo {author} {\bibfnamefont {B.}~\bibnamefont {Machielse}}, \bibinfo {author} {\bibfnamefont {E.~N.}\ \bibnamefont {Knall}}, \bibinfo {author} {\bibfnamefont {M.}~\bibnamefont {Sutula}}, \bibinfo {author} {\bibfnamefont {G.}~\bibnamefont {Baranes}}, \bibinfo {author} {\bibfnamefont {N.}~\bibnamefont {Sinclair}}, \bibinfo {author} {\bibfnamefont {C.}~\bibnamefont {De-Eknamkul}}, \bibinfo {author} {\bibfnamefont {D.~S.}\ \bibnamefont {Levonian}}, \bibinfo {author} {\bibfnamefont {M.~K.}\ \bibnamefont {Bhaskar}}, \bibinfo {author} {\bibfnamefont {H.}~\bibnamefont {Park}}, \bibinfo {author}
  {\bibfnamefont {M.}~\bibnamefont {Lončar}},\ and\ \bibinfo {author} {\bibfnamefont {M.~D.}\ \bibnamefont {Lukin}},\ }\href {http://dx.doi.org/10.1038/s41586-024-07252-z} {\bibfield  {journal} {\bibinfo  {journal} {Nature}\ }\textbf {\bibinfo {volume} {629}},\ \bibinfo {pages} {573–578} (\bibinfo {year} {2024})}\BibitemShut {NoStop}%
\bibitem [{\citenamefont {Bradley}\ \emph {et~al.}(2019)\citenamefont {Bradley}, \citenamefont {Randall}, \citenamefont {Abobeih}, \citenamefont {Berrevoets}, \citenamefont {Degen}, \citenamefont {Bakker}, \citenamefont {Markham}, \citenamefont {Twitchen},\ and\ \citenamefont {Taminiau}}]{Bradley_2019}%
  \BibitemOpen
  \bibfield  {author} {\bibinfo {author} {\bibfnamefont {C.~E.}\ \bibnamefont {Bradley}}, \bibinfo {author} {\bibfnamefont {J.}~\bibnamefont {Randall}}, \bibinfo {author} {\bibfnamefont {M.~H.}\ \bibnamefont {Abobeih}}, \bibinfo {author} {\bibfnamefont {R.~C.}\ \bibnamefont {Berrevoets}}, \bibinfo {author} {\bibfnamefont {M.~J.}\ \bibnamefont {Degen}}, \bibinfo {author} {\bibfnamefont {M.~A.}\ \bibnamefont {Bakker}}, \bibinfo {author} {\bibfnamefont {M.}~\bibnamefont {Markham}}, \bibinfo {author} {\bibfnamefont {D.~J.}\ \bibnamefont {Twitchen}},\ and\ \bibinfo {author} {\bibfnamefont {T.~H.}\ \bibnamefont {Taminiau}},\ }\href {http://dx.doi.org/10.1103/PhysRevX.9.031045} {\bibfield  {journal} {\bibinfo  {journal} {Physical Review X}\ }\textbf {\bibinfo {volume} {9}} (\bibinfo {year} {2019})}\BibitemShut {NoStop}%
\bibitem [{\citenamefont {Nagata}\ \emph {et~al.}(2018)\citenamefont {Nagata}, \citenamefont {Kuramitani}, \citenamefont {Sekiguchi},\ and\ \citenamefont {Kosaka}}]{Nagata_2018}%
  \BibitemOpen
  \bibfield  {author} {\bibinfo {author} {\bibfnamefont {K.}~\bibnamefont {Nagata}}, \bibinfo {author} {\bibfnamefont {K.}~\bibnamefont {Kuramitani}}, \bibinfo {author} {\bibfnamefont {Y.}~\bibnamefont {Sekiguchi}},\ and\ \bibinfo {author} {\bibfnamefont {H.}~\bibnamefont {Kosaka}},\ }\href {http://dx.doi.org/10.1038/s41467-018-05664-w} {\bibfield  {journal} {\bibinfo  {journal} {Nature Communications}\ }\textbf {\bibinfo {volume} {9}} (\bibinfo {year} {2018})}\BibitemShut {NoStop}%
\bibitem [{\citenamefont {Hermans}\ \emph {et~al.}(2022)\citenamefont {Hermans}, \citenamefont {Pompili}, \citenamefont {Beukers}, \citenamefont {Baier}, \citenamefont {Borregaard},\ and\ \citenamefont {Hanson}}]{Hermans_2022}%
  \BibitemOpen
  \bibfield  {author} {\bibinfo {author} {\bibfnamefont {S.~L.~N.}\ \bibnamefont {Hermans}}, \bibinfo {author} {\bibfnamefont {M.}~\bibnamefont {Pompili}}, \bibinfo {author} {\bibfnamefont {H.~K.~C.}\ \bibnamefont {Beukers}}, \bibinfo {author} {\bibfnamefont {S.}~\bibnamefont {Baier}}, \bibinfo {author} {\bibfnamefont {J.}~\bibnamefont {Borregaard}},\ and\ \bibinfo {author} {\bibfnamefont {R.}~\bibnamefont {Hanson}},\ }\href {http://dx.doi.org/10.1038/s41586-022-04697-y} {\bibfield  {journal} {\bibinfo  {journal} {Nature}\ }\textbf {\bibinfo {volume} {605}},\ \bibinfo {pages} {663} (\bibinfo {year} {2022})}\BibitemShut {NoStop}%
\bibitem [{\citenamefont {Bernien}\ \emph {et~al.}(2013)\citenamefont {Bernien}, \citenamefont {Hensen}, \citenamefont {Pfaff}, \citenamefont {Koolstra}, \citenamefont {Blok}, \citenamefont {Robledo}, \citenamefont {Taminiau}, \citenamefont {Markham}, \citenamefont {Twitchen}, \citenamefont {Childress},\ and\ \citenamefont {Hanson}}]{Bernien_2013}%
  \BibitemOpen
  \bibfield  {author} {\bibinfo {author} {\bibfnamefont {H.}~\bibnamefont {Bernien}}, \bibinfo {author} {\bibfnamefont {B.}~\bibnamefont {Hensen}}, \bibinfo {author} {\bibfnamefont {W.}~\bibnamefont {Pfaff}}, \bibinfo {author} {\bibfnamefont {G.}~\bibnamefont {Koolstra}}, \bibinfo {author} {\bibfnamefont {M.~S.}\ \bibnamefont {Blok}}, \bibinfo {author} {\bibfnamefont {L.}~\bibnamefont {Robledo}}, \bibinfo {author} {\bibfnamefont {T.~H.}\ \bibnamefont {Taminiau}}, \bibinfo {author} {\bibfnamefont {M.}~\bibnamefont {Markham}}, \bibinfo {author} {\bibfnamefont {D.~J.}\ \bibnamefont {Twitchen}}, \bibinfo {author} {\bibfnamefont {L.}~\bibnamefont {Childress}},\ and\ \bibinfo {author} {\bibfnamefont {R.}~\bibnamefont {Hanson}},\ }\href {http://dx.doi.org/10.1038/nature12016} {\bibfield  {journal} {\bibinfo  {journal} {Nature}\ }\textbf {\bibinfo {volume} {497}},\ \bibinfo {pages} {86} (\bibinfo {year} {2013})}\BibitemShut {NoStop}%
\bibitem [{\citenamefont {Barrett}\ and\ \citenamefont {Kok}(2005)}]{Barrett_2005}%
  \BibitemOpen
  \bibfield  {author} {\bibinfo {author} {\bibfnamefont {S.~D.}\ \bibnamefont {Barrett}}\ and\ \bibinfo {author} {\bibfnamefont {P.}~\bibnamefont {Kok}},\ }\href {http://dx.doi.org/10.1103/PhysRevA.71.060310} {\bibfield  {journal} {\bibinfo  {journal} {Physical Review A}\ }\textbf {\bibinfo {volume} {71}} (\bibinfo {year} {2005})}\BibitemShut {NoStop}%
\bibitem [{\citenamefont {Hensen}\ \emph {et~al.}(2015)\citenamefont {Hensen}, \citenamefont {Bernien}, \citenamefont {Dréau}, \citenamefont {Reiserer}, \citenamefont {Kalb}, \citenamefont {Blok}, \citenamefont {Ruitenberg}, \citenamefont {Vermeulen}, \citenamefont {Schouten}, \citenamefont {Abellán}, \citenamefont {Amaya}, \citenamefont {Pruneri}, \citenamefont {Mitchell}, \citenamefont {Markham}, \citenamefont {Twitchen}, \citenamefont {Elkouss}, \citenamefont {Wehner}, \citenamefont {Taminiau},\ and\ \citenamefont {Hanson}}]{Hensen_2015}%
  \BibitemOpen
  \bibfield  {author} {\bibinfo {author} {\bibfnamefont {B.}~\bibnamefont {Hensen}}, \bibinfo {author} {\bibfnamefont {H.}~\bibnamefont {Bernien}}, \bibinfo {author} {\bibfnamefont {A.~E.}\ \bibnamefont {Dréau}}, \bibinfo {author} {\bibfnamefont {A.}~\bibnamefont {Reiserer}}, \bibinfo {author} {\bibfnamefont {N.}~\bibnamefont {Kalb}}, \bibinfo {author} {\bibfnamefont {M.~S.}\ \bibnamefont {Blok}}, \bibinfo {author} {\bibfnamefont {J.}~\bibnamefont {Ruitenberg}}, \bibinfo {author} {\bibfnamefont {R.~F.~L.}\ \bibnamefont {Vermeulen}}, \bibinfo {author} {\bibfnamefont {R.~N.}\ \bibnamefont {Schouten}}, \bibinfo {author} {\bibfnamefont {C.}~\bibnamefont {Abellán}}, \bibinfo {author} {\bibfnamefont {W.}~\bibnamefont {Amaya}}, \bibinfo {author} {\bibfnamefont {V.}~\bibnamefont {Pruneri}}, \bibinfo {author} {\bibfnamefont {M.~W.}\ \bibnamefont {Mitchell}}, \bibinfo {author} {\bibfnamefont {M.}~\bibnamefont {Markham}}, \bibinfo {author} {\bibfnamefont {D.~J.}\ \bibnamefont {Twitchen}}, \bibinfo {author}
  {\bibfnamefont {D.}~\bibnamefont {Elkouss}}, \bibinfo {author} {\bibfnamefont {S.}~\bibnamefont {Wehner}}, \bibinfo {author} {\bibfnamefont {T.~H.}\ \bibnamefont {Taminiau}},\ and\ \bibinfo {author} {\bibfnamefont {R.}~\bibnamefont {Hanson}},\ }\href {http://dx.doi.org/10.1038/nature15759} {\bibfield  {journal} {\bibinfo  {journal} {Nature}\ }\textbf {\bibinfo {volume} {526}},\ \bibinfo {pages} {682–686} (\bibinfo {year} {2015})}\BibitemShut {NoStop}%
\bibitem [{\citenamefont {Cabrillo}\ \emph {et~al.}(1999)\citenamefont {Cabrillo}, \citenamefont {Cirac}, \citenamefont {García-Fernández},\ and\ \citenamefont {Zoller}}]{Cabrillo_1999}%
  \BibitemOpen
  \bibfield  {author} {\bibinfo {author} {\bibfnamefont {C.}~\bibnamefont {Cabrillo}}, \bibinfo {author} {\bibfnamefont {J.~I.}\ \bibnamefont {Cirac}}, \bibinfo {author} {\bibfnamefont {P.}~\bibnamefont {García-Fernández}},\ and\ \bibinfo {author} {\bibfnamefont {P.}~\bibnamefont {Zoller}},\ }\href {http://dx.doi.org/10.1103/PhysRevA.59.1025} {\bibfield  {journal} {\bibinfo  {journal} {Physical Review A}\ }\textbf {\bibinfo {volume} {59}},\ \bibinfo {pages} {1025–1033} (\bibinfo {year} {1999})}\BibitemShut {NoStop}%
\bibitem [{\citenamefont {Duan}\ \emph {et~al.}(2001)\citenamefont {Duan}, \citenamefont {Lukin}, \citenamefont {Cirac},\ and\ \citenamefont {Zoller}}]{Duan_2001}%
  \BibitemOpen
  \bibfield  {author} {\bibinfo {author} {\bibfnamefont {L.-M.}\ \bibnamefont {Duan}}, \bibinfo {author} {\bibfnamefont {M.~D.}\ \bibnamefont {Lukin}}, \bibinfo {author} {\bibfnamefont {J.~I.}\ \bibnamefont {Cirac}},\ and\ \bibinfo {author} {\bibfnamefont {P.}~\bibnamefont {Zoller}},\ }\href {http://dx.doi.org/10.1038/35106500} {\bibfield  {journal} {\bibinfo  {journal} {Nature}\ }\textbf {\bibinfo {volume} {414}},\ \bibinfo {pages} {413} (\bibinfo {year} {2001})}\BibitemShut {NoStop}%
\bibitem [{\citenamefont {Stolk}\ \emph {et~al.}(2024)\citenamefont {Stolk}, \citenamefont {van~der Enden}, \citenamefont {Slater}, \citenamefont {te~Raa-Derckx}, \citenamefont {Botma}, \citenamefont {van Rantwijk}, \citenamefont {Biemond}, \citenamefont {Hagen}, \citenamefont {Herfst}, \citenamefont {Koek}, \citenamefont {Meskers}, \citenamefont {Vollmer}, \citenamefont {van Zwet}, \citenamefont {Markham}, \citenamefont {Edmonds}, \citenamefont {Geus}, \citenamefont {Elsen}, \citenamefont {Jungbluth}, \citenamefont {Haefner}, \citenamefont {Tresp}, \citenamefont {Stuhler}, \citenamefont {Ritter},\ and\ \citenamefont {Hanson}}]{Stolk_2024}%
  \BibitemOpen
  \bibfield  {author} {\bibinfo {author} {\bibfnamefont {A.~J.}\ \bibnamefont {Stolk}}, \bibinfo {author} {\bibfnamefont {K.~L.}\ \bibnamefont {van~der Enden}}, \bibinfo {author} {\bibfnamefont {M.-C.}\ \bibnamefont {Slater}}, \bibinfo {author} {\bibfnamefont {I.}~\bibnamefont {te~Raa-Derckx}}, \bibinfo {author} {\bibfnamefont {P.}~\bibnamefont {Botma}}, \bibinfo {author} {\bibfnamefont {J.}~\bibnamefont {van Rantwijk}}, \bibinfo {author} {\bibfnamefont {J.~J.~B.}\ \bibnamefont {Biemond}}, \bibinfo {author} {\bibfnamefont {R.~A.~J.}\ \bibnamefont {Hagen}}, \bibinfo {author} {\bibfnamefont {R.~W.}\ \bibnamefont {Herfst}}, \bibinfo {author} {\bibfnamefont {W.~D.}\ \bibnamefont {Koek}}, \bibinfo {author} {\bibfnamefont {A.~J.~H.}\ \bibnamefont {Meskers}}, \bibinfo {author} {\bibfnamefont {R.}~\bibnamefont {Vollmer}}, \bibinfo {author} {\bibfnamefont {E.~J.}\ \bibnamefont {van Zwet}}, \bibinfo {author} {\bibfnamefont {M.}~\bibnamefont {Markham}}, \bibinfo {author} {\bibfnamefont {A.~M.}\ \bibnamefont {Edmonds}},
  \bibinfo {author} {\bibfnamefont {J.~F.}\ \bibnamefont {Geus}}, \bibinfo {author} {\bibfnamefont {F.}~\bibnamefont {Elsen}}, \bibinfo {author} {\bibfnamefont {B.}~\bibnamefont {Jungbluth}}, \bibinfo {author} {\bibfnamefont {C.}~\bibnamefont {Haefner}}, \bibinfo {author} {\bibfnamefont {C.}~\bibnamefont {Tresp}}, \bibinfo {author} {\bibfnamefont {J.}~\bibnamefont {Stuhler}}, \bibinfo {author} {\bibfnamefont {S.}~\bibnamefont {Ritter}},\ and\ \bibinfo {author} {\bibfnamefont {R.}~\bibnamefont {Hanson}},\ }\href {http://dx.doi.org/10.1126/sciadv.adp6442} {\bibfield  {journal} {\bibinfo  {journal} {Science Advances}\ }\textbf {\bibinfo {volume} {10}} (\bibinfo {year} {2024})}\BibitemShut {NoStop}%
\bibitem [{\citenamefont {Yang}\ \emph {et~al.}(2016)\citenamefont {Yang}, \citenamefont {Wang}, \citenamefont {Rao}, \citenamefont {Hien~Tran}, \citenamefont {Momenzadeh}, \citenamefont {Markham}, \citenamefont {Twitchen}, \citenamefont {Wang}, \citenamefont {Yang}, \citenamefont {Stöhr}, \citenamefont {Neumann}, \citenamefont {Kosaka},\ and\ \citenamefont {Wrachtrup}}]{Yang_2016}%
  \BibitemOpen
  \bibfield  {author} {\bibinfo {author} {\bibfnamefont {S.}~\bibnamefont {Yang}}, \bibinfo {author} {\bibfnamefont {Y.}~\bibnamefont {Wang}}, \bibinfo {author} {\bibfnamefont {D.~D.~B.}\ \bibnamefont {Rao}}, \bibinfo {author} {\bibfnamefont {T.}~\bibnamefont {Hien~Tran}}, \bibinfo {author} {\bibfnamefont {A.~S.}\ \bibnamefont {Momenzadeh}}, \bibinfo {author} {\bibfnamefont {M.}~\bibnamefont {Markham}}, \bibinfo {author} {\bibfnamefont {D.~J.}\ \bibnamefont {Twitchen}}, \bibinfo {author} {\bibfnamefont {P.}~\bibnamefont {Wang}}, \bibinfo {author} {\bibfnamefont {W.}~\bibnamefont {Yang}}, \bibinfo {author} {\bibfnamefont {R.}~\bibnamefont {Stöhr}}, \bibinfo {author} {\bibfnamefont {P.}~\bibnamefont {Neumann}}, \bibinfo {author} {\bibfnamefont {H.}~\bibnamefont {Kosaka}},\ and\ \bibinfo {author} {\bibfnamefont {J.}~\bibnamefont {Wrachtrup}},\ }\href {http://dx.doi.org/10.1038/nphoton.2016.103} {\bibfield  {journal} {\bibinfo  {journal} {Nature Photonics}\ }\textbf {\bibinfo {volume} {10}},\ \bibinfo {pages}
  {507–511} (\bibinfo {year} {2016})}\BibitemShut {NoStop}%
\bibitem [{\citenamefont {Tsurumoto}\ \emph {et~al.}(2019)\citenamefont {Tsurumoto}, \citenamefont {Kuroiwa}, \citenamefont {Kano}, \citenamefont {Sekiguchi},\ and\ \citenamefont {Kosaka}}]{Tsurumoto_2019}%
  \BibitemOpen
  \bibfield  {author} {\bibinfo {author} {\bibfnamefont {K.}~\bibnamefont {Tsurumoto}}, \bibinfo {author} {\bibfnamefont {R.}~\bibnamefont {Kuroiwa}}, \bibinfo {author} {\bibfnamefont {H.}~\bibnamefont {Kano}}, \bibinfo {author} {\bibfnamefont {Y.}~\bibnamefont {Sekiguchi}},\ and\ \bibinfo {author} {\bibfnamefont {H.}~\bibnamefont {Kosaka}},\ }\href {http://dx.doi.org/10.1038/s42005-019-0158-0} {\bibfield  {journal} {\bibinfo  {journal} {Communications Physics}\ }\textbf {\bibinfo {volume} {2}} (\bibinfo {year} {2019})}\BibitemShut {NoStop}%
\bibitem [{\citenamefont {Maze}\ \emph {et~al.}(2011)\citenamefont {Maze}, \citenamefont {Gali}, \citenamefont {Togan}, \citenamefont {Chu}, \citenamefont {Trifonov}, \citenamefont {Kaxiras},\ and\ \citenamefont {Lukin}}]{Maze_2011}%
  \BibitemOpen
  \bibfield  {author} {\bibinfo {author} {\bibfnamefont {J.~R.}\ \bibnamefont {Maze}}, \bibinfo {author} {\bibfnamefont {A.}~\bibnamefont {Gali}}, \bibinfo {author} {\bibfnamefont {E.}~\bibnamefont {Togan}}, \bibinfo {author} {\bibfnamefont {Y.}~\bibnamefont {Chu}}, \bibinfo {author} {\bibfnamefont {A.}~\bibnamefont {Trifonov}}, \bibinfo {author} {\bibfnamefont {E.}~\bibnamefont {Kaxiras}},\ and\ \bibinfo {author} {\bibfnamefont {M.~D.}\ \bibnamefont {Lukin}},\ }\href {http://dx.doi.org/10.1088/1367-2630/13/2/025025} {\bibfield  {journal} {\bibinfo  {journal} {New Journal of Physics}\ }\textbf {\bibinfo {volume} {13}},\ \bibinfo {pages} {025025} (\bibinfo {year} {2011})}\BibitemShut {NoStop}%
\bibitem [{\citenamefont {Kosaka}\ and\ \citenamefont {Niikura}(2015)}]{Kosaka_2015}%
  \BibitemOpen
  \bibfield  {author} {\bibinfo {author} {\bibfnamefont {H.}~\bibnamefont {Kosaka}}\ and\ \bibinfo {author} {\bibfnamefont {N.}~\bibnamefont {Niikura}},\ }\bibfield  {journal} {\bibinfo  {journal} {Physical Review Letters}\ }\textbf {\bibinfo {volume} {114}},\ \href {https://doi.org/10.1103/physrevlett.114.053603} {10.1103/physrevlett.114.053603} (\bibinfo {year} {2015})\BibitemShut {NoStop}%
\bibitem [{\citenamefont {Robledo}\ \emph {et~al.}(2011)\citenamefont {Robledo}, \citenamefont {Childress}, \citenamefont {Bernien}, \citenamefont {Hensen}, \citenamefont {Alkemade},\ and\ \citenamefont {Hanson}}]{Robledo_2011}%
  \BibitemOpen
  \bibfield  {author} {\bibinfo {author} {\bibfnamefont {L.}~\bibnamefont {Robledo}}, \bibinfo {author} {\bibfnamefont {L.}~\bibnamefont {Childress}}, \bibinfo {author} {\bibfnamefont {H.}~\bibnamefont {Bernien}}, \bibinfo {author} {\bibfnamefont {B.}~\bibnamefont {Hensen}}, \bibinfo {author} {\bibfnamefont {P.~F.~A.}\ \bibnamefont {Alkemade}},\ and\ \bibinfo {author} {\bibfnamefont {R.}~\bibnamefont {Hanson}},\ }\href {http://dx.doi.org/10.1038/nature10401} {\bibfield  {journal} {\bibinfo  {journal} {Nature}\ }\textbf {\bibinfo {volume} {477}},\ \bibinfo {pages} {574} (\bibinfo {year} {2011})}\BibitemShut {NoStop}%
\bibitem [{\citenamefont {Khaneja}\ \emph {et~al.}(2005)\citenamefont {Khaneja}, \citenamefont {Reiss}, \citenamefont {Kehlet}, \citenamefont {Schulte-Herbrüggen},\ and\ \citenamefont {Glaser}}]{Khaneja_2005}%
  \BibitemOpen
  \bibfield  {author} {\bibinfo {author} {\bibfnamefont {N.}~\bibnamefont {Khaneja}}, \bibinfo {author} {\bibfnamefont {T.}~\bibnamefont {Reiss}}, \bibinfo {author} {\bibfnamefont {C.}~\bibnamefont {Kehlet}}, \bibinfo {author} {\bibfnamefont {T.}~\bibnamefont {Schulte-Herbrüggen}},\ and\ \bibinfo {author} {\bibfnamefont {S.~J.}\ \bibnamefont {Glaser}},\ }\href {http://dx.doi.org/10.1016/j.jmr.2004.11.004} {\bibfield  {journal} {\bibinfo  {journal} {Journal of Magnetic Resonance}\ }\textbf {\bibinfo {volume} {172}},\ \bibinfo {pages} {296–305} (\bibinfo {year} {2005})}\BibitemShut {NoStop}%
\bibitem [{\citenamefont {Kamimaki}\ \emph {et~al.}(2023)\citenamefont {Kamimaki}, \citenamefont {Wakamatsu}, \citenamefont {Mikata}, \citenamefont {Sekiguchi},\ and\ \citenamefont {Kosaka}}]{kamimaki_2023}%
  \BibitemOpen
  \bibfield  {author} {\bibinfo {author} {\bibfnamefont {A.}~\bibnamefont {Kamimaki}}, \bibinfo {author} {\bibfnamefont {K.}~\bibnamefont {Wakamatsu}}, \bibinfo {author} {\bibfnamefont {K.}~\bibnamefont {Mikata}}, \bibinfo {author} {\bibfnamefont {Y.}~\bibnamefont {Sekiguchi}},\ and\ \bibinfo {author} {\bibfnamefont {H.}~\bibnamefont {Kosaka}},\ }\href {http://dx.doi.org/10.1038/s41534-023-00771-z} {\bibfield  {journal} {\bibinfo  {journal} {npj Quantum Information}\ }\textbf {\bibinfo {volume} {9}} (\bibinfo {year} {2023})}\BibitemShut {NoStop}%
\bibitem [{\citenamefont {Batalov}\ \emph {et~al.}(2009)\citenamefont {Batalov}, \citenamefont {Jacques}, \citenamefont {Kaiser}, \citenamefont {Siyushev}, \citenamefont {Neumann}, \citenamefont {Rogers}, \citenamefont {McMurtrie}, \citenamefont {Manson}, \citenamefont {Jelezko},\ and\ \citenamefont {Wrachtrup}}]{Batalov_2009}%
  \BibitemOpen
  \bibfield  {author} {\bibinfo {author} {\bibfnamefont {A.}~\bibnamefont {Batalov}}, \bibinfo {author} {\bibfnamefont {V.}~\bibnamefont {Jacques}}, \bibinfo {author} {\bibfnamefont {F.}~\bibnamefont {Kaiser}}, \bibinfo {author} {\bibfnamefont {P.}~\bibnamefont {Siyushev}}, \bibinfo {author} {\bibfnamefont {P.}~\bibnamefont {Neumann}}, \bibinfo {author} {\bibfnamefont {L.~J.}\ \bibnamefont {Rogers}}, \bibinfo {author} {\bibfnamefont {R.~L.}\ \bibnamefont {McMurtrie}}, \bibinfo {author} {\bibfnamefont {N.~B.}\ \bibnamefont {Manson}}, \bibinfo {author} {\bibfnamefont {F.}~\bibnamefont {Jelezko}},\ and\ \bibinfo {author} {\bibfnamefont {J.}~\bibnamefont {Wrachtrup}},\ }\href {http://dx.doi.org/10.1103/PhysRevLett.102.195506} {\bibfield  {journal} {\bibinfo  {journal} {Physical Review Letters}\ }\textbf {\bibinfo {volume} {102}} (\bibinfo {year} {2009})}\BibitemShut {NoStop}%
\bibitem [{\citenamefont {Bersin}\ \emph {et~al.}(2024)\citenamefont {Bersin}, \citenamefont {Grein}, \citenamefont {Sutula}, \citenamefont {Murphy}, \citenamefont {Huan}, \citenamefont {Stevens}, \citenamefont {Suleymanzade}, \citenamefont {Lee}, \citenamefont {Riedinger}, \citenamefont {Starling}, \citenamefont {Stas}, \citenamefont {Knaut}, \citenamefont {Sinclair}, \citenamefont {Assumpcao}, \citenamefont {Wei}, \citenamefont {Knall}, \citenamefont {Machielse}, \citenamefont {Sukachev}, \citenamefont {Levonian}, \citenamefont {Bhaskar}, \citenamefont {Lon{\v{c}}~ar}, \citenamefont {Hamilton}, \citenamefont {Lukin}, \citenamefont {Englund},\ and\ \citenamefont {Dixon}}]{Bersin_2024}%
  \BibitemOpen
  \bibfield  {author} {\bibinfo {author} {\bibfnamefont {E.}~\bibnamefont {Bersin}}, \bibinfo {author} {\bibfnamefont {M.}~\bibnamefont {Grein}}, \bibinfo {author} {\bibfnamefont {M.}~\bibnamefont {Sutula}}, \bibinfo {author} {\bibfnamefont {R.}~\bibnamefont {Murphy}}, \bibinfo {author} {\bibfnamefont {Y.~Q.}\ \bibnamefont {Huan}}, \bibinfo {author} {\bibfnamefont {M.}~\bibnamefont {Stevens}}, \bibinfo {author} {\bibfnamefont {A.}~\bibnamefont {Suleymanzade}}, \bibinfo {author} {\bibfnamefont {C.}~\bibnamefont {Lee}}, \bibinfo {author} {\bibfnamefont {R.}~\bibnamefont {Riedinger}}, \bibinfo {author} {\bibfnamefont {D.~J.}\ \bibnamefont {Starling}}, \bibinfo {author} {\bibfnamefont {P.-J.}\ \bibnamefont {Stas}}, \bibinfo {author} {\bibfnamefont {C.~M.}\ \bibnamefont {Knaut}}, \bibinfo {author} {\bibfnamefont {N.}~\bibnamefont {Sinclair}}, \bibinfo {author} {\bibfnamefont {D.~R.}\ \bibnamefont {Assumpcao}}, \bibinfo {author} {\bibfnamefont {Y.-C.}\ \bibnamefont {Wei}}, \bibinfo {author} {\bibfnamefont {E.~N.}\
  \bibnamefont {Knall}}, \bibinfo {author} {\bibfnamefont {B.}~\bibnamefont {Machielse}}, \bibinfo {author} {\bibfnamefont {D.~D.}\ \bibnamefont {Sukachev}}, \bibinfo {author} {\bibfnamefont {D.~S.}\ \bibnamefont {Levonian}}, \bibinfo {author} {\bibfnamefont {M.~K.}\ \bibnamefont {Bhaskar}}, \bibinfo {author} {\bibfnamefont {M.}~\bibnamefont {Lon{\v{c}}~ar}}, \bibinfo {author} {\bibfnamefont {S.}~\bibnamefont {Hamilton}}, \bibinfo {author} {\bibfnamefont {M.}~\bibnamefont {Lukin}}, \bibinfo {author} {\bibfnamefont {D.}~\bibnamefont {Englund}},\ and\ \bibinfo {author} {\bibfnamefont {P.~B.}\ \bibnamefont {Dixon}},\ }\href {http://dx.doi.org/10.1103/PhysRevApplied.21.014024} {\bibfield  {journal} {\bibinfo  {journal} {Physical Review Applied}\ }\textbf {\bibinfo {volume} {21}} (\bibinfo {year} {2024})}\BibitemShut {NoStop}%
\bibitem [{\citenamefont {McCullian}\ \emph {et~al.}(2022)\citenamefont {McCullian}, \citenamefont {Cheung}, \citenamefont {Chen},\ and\ \citenamefont {Fuchs}}]{McCullian_2022}%
  \BibitemOpen
  \bibfield  {author} {\bibinfo {author} {\bibfnamefont {B.}~\bibnamefont {McCullian}}, \bibinfo {author} {\bibfnamefont {H.}~\bibnamefont {Cheung}}, \bibinfo {author} {\bibfnamefont {H.}~\bibnamefont {Chen}},\ and\ \bibinfo {author} {\bibfnamefont {G.}~\bibnamefont {Fuchs}},\ }\href {http://dx.doi.org/10.1103/PhysRevApplied.18.064011} {\bibfield  {journal} {\bibinfo  {journal} {Physical Review Applied}\ }\textbf {\bibinfo {volume} {18}} (\bibinfo {year} {2022})}\BibitemShut {NoStop}%
\bibitem [{\citenamefont {Sekiguchi}\ \emph {et~al.}(2021)\citenamefont {Sekiguchi}, \citenamefont {Yasui}, \citenamefont {Tsurumoto}, \citenamefont {Koga}, \citenamefont {Reyes},\ and\ \citenamefont {Kosaka}}]{Sekiguchi_2021}%
  \BibitemOpen
  \bibfield  {author} {\bibinfo {author} {\bibfnamefont {Y.}~\bibnamefont {Sekiguchi}}, \bibinfo {author} {\bibfnamefont {Y.}~\bibnamefont {Yasui}}, \bibinfo {author} {\bibfnamefont {K.}~\bibnamefont {Tsurumoto}}, \bibinfo {author} {\bibfnamefont {Y.}~\bibnamefont {Koga}}, \bibinfo {author} {\bibfnamefont {R.}~\bibnamefont {Reyes}},\ and\ \bibinfo {author} {\bibfnamefont {H.}~\bibnamefont {Kosaka}},\ }\href {http://dx.doi.org/10.1038/s42005-021-00767-1} {\bibfield  {journal} {\bibinfo  {journal} {Communications Physics}\ }\textbf {\bibinfo {volume} {4}} (\bibinfo {year} {2021})}\BibitemShut {NoStop}%
\bibitem [{\citenamefont {Bassett}\ \emph {et~al.}(2011)\citenamefont {Bassett}, \citenamefont {Heremans}, \citenamefont {Yale}, \citenamefont {Buckley},\ and\ \citenamefont {Awschalom}}]{Bassett_2011}%
  \BibitemOpen
  \bibfield  {author} {\bibinfo {author} {\bibfnamefont {L.~C.}\ \bibnamefont {Bassett}}, \bibinfo {author} {\bibfnamefont {F.~J.}\ \bibnamefont {Heremans}}, \bibinfo {author} {\bibfnamefont {C.~G.}\ \bibnamefont {Yale}}, \bibinfo {author} {\bibfnamefont {B.~B.}\ \bibnamefont {Buckley}},\ and\ \bibinfo {author} {\bibfnamefont {D.~D.}\ \bibnamefont {Awschalom}},\ }\href {http://dx.doi.org/10.1103/PhysRevLett.107.266403} {\bibfield  {journal} {\bibinfo  {journal} {Physical Review Letters}\ }\textbf {\bibinfo {volume} {107}} (\bibinfo {year} {2011})}\BibitemShut {NoStop}%
\bibitem [{\citenamefont {Sangouard}\ \emph {et~al.}(2011)\citenamefont {Sangouard}, \citenamefont {Simon}, \citenamefont {de~Riedmatten},\ and\ \citenamefont {Gisin}}]{Sangouard_2011}%
  \BibitemOpen
  \bibfield  {author} {\bibinfo {author} {\bibfnamefont {N.}~\bibnamefont {Sangouard}}, \bibinfo {author} {\bibfnamefont {C.}~\bibnamefont {Simon}}, \bibinfo {author} {\bibfnamefont {H.}~\bibnamefont {de~Riedmatten}},\ and\ \bibinfo {author} {\bibfnamefont {N.}~\bibnamefont {Gisin}},\ }\href {http://dx.doi.org/10.1103/RevModPhys.83.33} {\bibfield  {journal} {\bibinfo  {journal} {Reviews of Modern Physics}\ }\textbf {\bibinfo {volume} {83}},\ \bibinfo {pages} {33} (\bibinfo {year} {2011})}\BibitemShut {NoStop}%
\bibitem [{\citenamefont {Jung}\ \emph {et~al.}(2019)\citenamefont {Jung}, \citenamefont {Görlitz}, \citenamefont {Kambs}, \citenamefont {Pauly}, \citenamefont {Raatz}, \citenamefont {Nelz}, \citenamefont {Neu}, \citenamefont {Edmonds}, \citenamefont {Markham}, \citenamefont {Mücklich}, \citenamefont {Meijer},\ and\ \citenamefont {Becher}}]{Jung_2019}%
  \BibitemOpen
  \bibfield  {author} {\bibinfo {author} {\bibfnamefont {T.}~\bibnamefont {Jung}}, \bibinfo {author} {\bibfnamefont {J.}~\bibnamefont {Görlitz}}, \bibinfo {author} {\bibfnamefont {B.}~\bibnamefont {Kambs}}, \bibinfo {author} {\bibfnamefont {C.}~\bibnamefont {Pauly}}, \bibinfo {author} {\bibfnamefont {N.}~\bibnamefont {Raatz}}, \bibinfo {author} {\bibfnamefont {R.}~\bibnamefont {Nelz}}, \bibinfo {author} {\bibfnamefont {E.}~\bibnamefont {Neu}}, \bibinfo {author} {\bibfnamefont {A.~M.}\ \bibnamefont {Edmonds}}, \bibinfo {author} {\bibfnamefont {M.}~\bibnamefont {Markham}}, \bibinfo {author} {\bibfnamefont {F.}~\bibnamefont {Mücklich}}, \bibinfo {author} {\bibfnamefont {J.}~\bibnamefont {Meijer}},\ and\ \bibinfo {author} {\bibfnamefont {C.}~\bibnamefont {Becher}},\ }\href {http://dx.doi.org/10.1063/1.5120120} {\bibfield  {journal} {\bibinfo  {journal} {APL Photonics}\ }\textbf {\bibinfo {volume} {4}} (\bibinfo {year} {2019})}\BibitemShut {NoStop}%
\end{thebibliography}
%

\end{document}